\documentclass{moriond}

\def\be{\begin{equation}}
\def\ee{\end{equation}}
\def\bea{\begin{eqnarray}}
\def\eea{\end{eqnarray}}

\newcommand{\Photo}{\includegraphics[height=35mm]{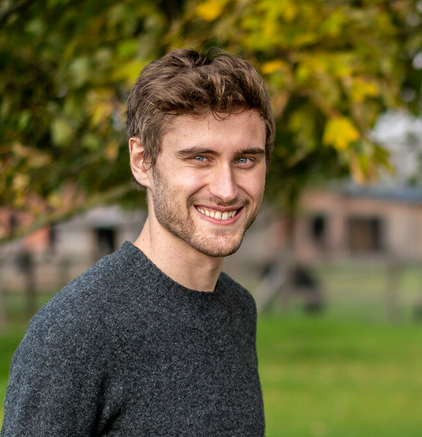}}

\begin{document}
\vspace*{4cm}
\title{Prospects for an isotropic gravitational wave background detection with Earth-based interferometric detectors and the threat of correlated noise}

\author{ K. Janssens }
\address{Universiteit Antwerpen, Prinsstraat 13, 2000 Antwerpen, Belgium \newline Universit\'e C$\hat{o}$te d’Azur, Observatoire C$\hat{o}$te d’Azur, CNRS, Artemis, 06304 Nice, France}

\maketitle\abstracts{ In this overview we discuss the prospects for a first detection of an isotropic gravitational wave background with earth-based interferometric detectors. Furthermore, we focus on how correlated noise sources could endanger such a detection with current generation of detectors. Finally, we project how correlated noise could significantly impede the potential of the future detector, the Einstein Telescope, in its search for an isotropic gravitational wave background. The triangular configuration of three (almost) co-located detectors, makes the Einstein Telescope especially prone to correlated noise sources.}

\section{Introduction}

After the first direct observation of gravitational waves (GWs) \cite{PhysRevLett.116.061102} by LIGO \cite{ADV_LIGO_2015} in 2015, the LIGO, Virgo \cite{VIRGO:2014yos} and KAGRA\cite{PhysRevD.88.043007} (LVK) collaborations have observed many more events in subsequent years. Currently several tens of binary black hole mergers are observed \cite{GWTC1,PhysRevX.11.021053,GWTC2.1,theligoscientificcollaboration2022population} as well as a handful of binary neutron star \cite{PhysRevLett.116.061102,PhysRevLett.119.161101,ApJL_892_L3} and neutron star black hole mergers \cite{Abbott_2021_NS}.

Whereas the search for more compact binary coalescence's (CBCs) continues, the LVK collaborations are also searching for other GW signals, such as the (stochastic) gravitational wave background ((S)GWB) \cite{Christensen_2018,PhysRevD.104.022004}. This is a background consisting of the incoherent superposition of many week GW sources and could be sourced by a wide variety of astrophysical and/or cosmological sources \cite{Christensen_2018,VANREMORTEL2022104003,Renzini:2022alw}.
Due to its low signal amplitude the GWB is typically hidden below the noise of individual detectors. Therefore one relies on correlating the data from two or more detectors with each other to `dig below' the noise-floor of individual detectors. If one assumes the detector noises are uncorrelated between two detectors, the only remaining signal in the correlated data is a potential GWB signal. In this work we will focus on the different noise sources that violate this assumption.

\section{Prospects for the detection of an isotropic GWB with Earth-based interferometric detectors}

The GWB we are guaranteed to observe with Earth-based detectors is the astrophysical background originating from distant CBC mergers which are to weak to be individually resolved. Current predictions indicate this signal should become observable after several years of data taking with LIGO and Virgo at their desgin A+/AdV+ sensitivity \cite{PhysRevD.104.022004}. This sensitivity is planned to be reached at the end of their fith observing run (O5), scheduled for the end of this decade \cite{LVKRunPlans}.
However, there are also many other possible GWB signals that could be observable at those sensitivities or of which the parameter space could be constrained in the case of a non-detection.

We want to point out that several more sophisticated analysis methods are being investigated to search for a GWB of intermittent nature as is expected for the GWB from binary black holes, such as the Stochastic Search for Intermittent GWB (SSI)\cite{Lawrence_2023SSI} and The Bayesian Search (TBS)\cite{PhysRevX.8.021019,PhysRevResearch.3.043049}. Even though both methods look very promising, currently they only have been tested on idealized Gaussian data and further investigations have to be performed how the pipelines react to more realistic data complications such as noise transients and correlated noise.

\section{Correlated noise and its impact on the search for an isotropic gravitational wave background}

We will start in Sec. \ref{sec:2G} with discussing the effect of correlated noise on the search for an isotropic GWB, with the current, second generation (2G) detectors, i.e. LIGO, Virgo and KAGRA. Afterwards, in Sec. \ref{sec:3G} we will discuss the impact on the European proposal for a third generation (3G) Earth based interferometric detector, the Einstein Telscope. Note that we will not be discussing the effect of correlated noise on the US based proposed detector Cosmic Explorer as the current literature is lacking in this domain.

\subsection{2G detectors}
\label{sec:2G}

With a distance of at least several thousands of kilometers between different detectors most of the noise sources are uncorrelated between the different 2G detectors. 
Here we will discuss the only broadband, environmental noise source that is known to potentially couple coherently to distant Earth-based interferometric detectors, i.e. magnetic field fluctuations.

\subsubsection{Correlations in magnetic field fluctuations}

Lightning strikes all around the world produce standing electromagnetic waves in the Earth surface - ionosphere cavity, known as the Schumann resonances\cite{Schumann1,Schumann2}. The fundamental mode has a frequency of 7.8Hz, and amplitude of $\sim$ 1 pT. Higher harmonics occur at 14Hz, 21Hz, ...
More recently also a 'stochastic' background due to the superposition of individually correlated lightning strikes was described and leads to an excess of magnetic correlation between 100Hz and 1kHz \cite{PhysRevD.107.022004}.
To understand to which level these magnetic fields couple to the detectors, the so called magnetic coupling function is measured for each detector. With measurements of environmental magnetic field fluctuations at each detector site and the magnetic coupling one can predict the effect of magnetic noise on each detectors sensitivity as well as on the search for a GWB.

This magnetic noise projection indicates there was no contamination from correlated magnetic noise during LVKs last observing run, O3. However, in case LIGO and Virgo will have the same magnetic coupling during their future observing runs, the search for an isotropic GWB might experience contamination from correlated noise when they reach A+/AdV+ sensitivity \cite{PhysRevD.107.022004}. More specifically correlations in magnetic field fluctuations could impact searches below 45Hz or above 150Hz, as shown in Fig. 10 of \cite{PhysRevD.107.022004}.

\subsection{3G detectors}
\label{sec:3G}

The European proposal for a next generation Earth-based GW detector, the Einstein Telescope (ET), uses a set-up of three co-located detectors in an equilateral triangular configuration. Due to the close proximity between the different detectors, also other noise sources can couple coherently between the different detectors. Here we will only discuss fundamental, environmental noise sources and their potential to limit the sensitivity for the search for an isotropic GWB. However, the detectors should be designed very carefully to not introduce additional correlated noise from e.g. the detectors infrastructure. 

\subsubsection{Correlations in magnetic field fluctuations}

Similarly as 2G detectors the ET could also be affected by correlations in magnetic field fluctuations. However due to its superior sensitivity, the potential contamination is also larger. As shown in a recent study \cite{PhysRevD.104.122006}, below $\sim$ 30Hz, the magnetic coupling of the ET, should be up to 4 orders of magnitude smaller, compared to the average magnetic coupling currently observed at LIGO and Virgo. In case this reduction in magnetic coupling function is not achieved, the search for an isotropic GWB would be highly contaminated by correlated noise in the low frequency regime \cite{PhysRevD.104.122006}.

\subsubsection{Correlations in seismic and Newtonian noise}

In the current configuration for the ET, multiple mirrors of two different detectors are separated by several hundreds of meters from each other, leading to potential coupling of correlated noise sources with coherence lengths on these distance scales.
A recent study \cite{PhysRevD.106.042008} investigates the levels of correlated seismic noise on these distance scales both at the surface and at underground facilities. With the planned levels of seismic isolation for the ET, the direct effect of correlations in the seismic noise is predicted to be limited and only affect the search for an isotropic GWB up to $\sim$ 5Hz \cite{PhysRevD.106.042008}. However this seismic noise also induces changes in the local gravitational field which leads to so-called gravity gradient or Newtonian noise. This noise exerts a direct force on the test masses for the ET and cannot be shielded for. With the level of seismic correlations observed on a distance scale of several hundreds of meters at the Homestake mine in the US, this leads to a Newtonian noise budget from seismic body waves as shown in Fig. 16 of \cite{PhysRevD.106.042008}. At 3Hz, the levels of correlated Newtonian noise are up to six orders of magnitude (90\% percentile) above the targeted sensitivity. The search for an isotropic GWB would be impacted by this noise source up to 40Hz, and provide no significant improvements on earlier results by searches at LVK A+ design sensitivity between $\sim$ 10Hz and 30Hz \cite{PhysRevD.106.042008}. The expected levels of noise subtraction are estimated to be $\sim$ 3-10 (realistic-optimistic), per detector, i.e. 9-100 per detector pair for GWB searchers\cite{PhysRevD.106.042008}. However even optimistic levels of noise subtraction do not suffice to prevent significant noise contamination from correlated Newtonian noise.

\section{Methods to deal with correlated noise}

Without going into any detail we highlight a couple of methods that have been proposed in the past to deal with correlated noise. A first tool, called GW-Geodesy \cite{GWGeodesy,PhysRevD.105.082001}, provides a method to identify the probability an observed GWB signal is from GW or noise origin.  
Secondly one could use Bayesian methods to include correlated noise terms in your data when performing parameter estimation \cite{BayesianGWMag}. Finally, one could consider to perform noise subtraction \cite{Coughlin:2018tjc,Thrane:2014yza,Coughlin:2016vor,Thrane:2013npa,PhysRevD.86.102001,PhysRevD.92.022001,Coughlin_2014_seismic,Coughlin_2016,PhysRevLett.121.221104,Tringali_2019,Badaracco_2019,Badaracco_2020,NN_Sardinia2020,10.1785/0220200186,Bader_2022,Koley_2022}, e.g. Wiener filtering.

\section{Conclusion}

We have shown that correlations in magnetic field fluctuations could impact the search for an isotropic gravitational wave background with LIGO, Virgo and KAGRA when they reach their A+ design sensitivity. The European proposal for the next generation detector, the Einstein Telescope, has a triangular configuration consisting of three detectors. Due to its increased sensitivity as well as the close proximity between the different detectors, the Einstein Telescope is much more prone to correlated noise. Environmental magnetic fields threaten to impact ETs search for a GWB up to $\sim$ 30 Hz, by several orders of magnitude at the lowest frequencies. At the same time correlated Newtonian noise from body waves could contaminate the search for a GWB up to $\sim$ 40 Hz, by up to six orders of magnitude at 3Hz.
Current mitigation techniques are lacking to entirely mitigate the projected effects of correlated magnetic and Newtonian noise for the ET and more research is needed to develop additional methods. Another solution would be to abandon the triangular detector configuration as is discussed in a recent study \cite{Branchesi:2023mws}.

\section*{Acknowledgments}

K.J. is supported by FWO-Vlaanderen via grant number 11C5720N.

\section*{References}

\bibliography{references}

\end{document}